# Robust ferromagnetism in highly strained SrCoO$_3$ thin films

Yujia Wang[1,2], Qing He[3*], Wenmei Ming[4], Mao-Hua Du[4], Nianpeng Lu[1,5], Clodomiro Cafolla[3], Jun Fujioka[6], Qinghua Zhang[5,7], Ding Zhang[1,2], Shengchun Shen[1,2], Yingjie Lyu[1,2], Alpha T. N'Diaye[8], Elke Arenholz[8], Lin Gu[5], Cewen Nan[7], Yoshinori Tokura[9,10], Satoshi Okamoto[4*] and Pu Yu[1,2,9*]

[1]State Key Laboratory of Low Dimensional Quantum Physics and Department of Physics, Tsinghua University, Beijing, 100084, China
[2]Frontier Science Center for Quantum Information, Beijing, 100084, China
[3]Department of Physics, Durham University, Durham DH1 3LE, UK
[4]Materials Science and Technology Division, Oak Ridge National Laboratory, Oak Ridge, Tennessee 37831, USA
[5]Beijing National Laboratory for Condensed Matter Physics, Institute of Physics, Chinese Academy of Science, Beijing 100190, China
[6]Graduate school of Pure and Applied Science, University of Tsukuba, Tsukuba, Ibaraki, Japan
[7]State Key Laboratory of New Ceramics and Fine Processing, School of Materials Science and Engineering, Tsinghua University, Beijing 100084, China
[8]Advanced Light Source, Lawrence Berkeley National Laboratory, Berkeley, California 94720, USA
[9]RIKEN Center for Emergent Matter Science (CEMS), Wako, Saitama 351-0198, Japan
[10]Tokyo College and Department of Applied Physics, University of Tokyo, Tokyo 113-8656, Japan
*Email: qing.he@durham.ac.uk, okapon@ornl.gov and yupu@mail.tsinghua.edu.cn

**Epitaxial strain provides important pathways to control the magnetic and electronic states in transition metal oxides. However, the large strain is usually accompanied by a strong reduction of the oxygen vacancy formation energy, which hinders the direct manipulation of their intrinsic properties. Here using a post-deposition ozone annealing method, we obtained a series of oxygen stoichiometric SrCoO$_3$ thin films with the tensile strain up to 3.0%. We observed a robust ferromagnetic ground state in all strained thin films, while interestingly the tensile strain triggers a distinct metal to insulator transition along with the increase of the tensile strain. The persistent ferromagnetic state across the electrical transition therefore suggests that the magnetic state is directly correlated with the localized electrons, rather than the itinerant ones, which then calls for further investigation of the intrinsic mechanism of this magnetic compound beyond the double-exchange mechanism.**

# I. INTRODUCTION

In complex oxides, the epitaxial strain in thin film structures provides an essential control parameter to manipulate their corresponding electronic and magnetic ground states due to the interplay among lattice, charge, orbital and spin degrees of freedom. A large group of transition-metal oxide materials have demonstrated the emergence of novel properties through strain engineering, such as enhanced transition temperatures in ferroelectric [1], ferromagnetic (FM) [2] and superconducting [3] orders, as well as emergent exotic electronic states [4-6], as widely reported in high quality epitaxially strained thin films.

Among such oxide materials family, particular attention has been payed to perovskite $SrCoO_3$ recently due to its intriguing FM metallic ground state [7,8] in the bulk compound attributed to the double-exchange mechanism proposed by Zener [9]. Here, according to the density functional theory (DFT), the Co $d^6$ ion antiferromagnetically couples with the ligand hole in the surrounding oxygen $p$ states through the $p$-$d$ hybridization to form the intermediate spin Co $d^6L$ state [10,11]. More sophisticated treatment of the correlation effects, using the DFT + dynamical mean field theory (LDA+DMFT) [12], claimed that the local moment in $SrCoO_3$ results from the coherent superposition of different atomic states rather than the intermediate spin state [13]. Motivated by the possible multiferroicity in strained transition-metal oxides [14], the magnetic ground state of strained $SrCoO_3$ was studied using the standard DFT. It was predicted that the ground state changes from a FM metallic state to an antiferromagnetic (AFM) metallic state, with a moderate tensile strain of 2.0% [15,16]. These theoretical predictions triggered a series of experimental studies to examine epitaxially strained $SrCoO_3$ thin films [17-20]. While most groups have obtained the FM metallic state in $SrCoO_3$ at low strain states, results on highly tensile-strained samples showed large diversity, which is likely due to the lack of good oxygen stoichiometry of the material resulting from the reduction of the oxygen vacancy formation energy due to large tensile strains [17,18,21]. Thus, the intrinsic magnetic and electronic states as well as the underlying mechanism of highly strained $SrCoO_3$ still remain elusive.

In this paper, we report our magnetic study of highly strained SrCoO$_3$ thin films (up to tensile strain of ~3.0%) with excellent oxygen stoichiometry, which is achieved by a new post-deposition ozone annealing process. To our surprise, the robust FM ground state is found in all tensile strained thin films, which is in stark contrast to previous theoretical predictions and experimental reports claiming a transition to AFM state. Interestingly, we observed a metal to insulator transition along with the increase of the tensile strains. To gain insight into this observation, we carried out DFT+DMFT calculations and found that the orbital occupation changes more strongly with strain than what DFT predicts. Based on this, we propose a super-exchange mechanism with alternating orbital ordering, which is consistent with the robust FM ground state with insulating transport.

## II. RESULTS

### A. Fabrication and characterization of highly strained SrCoO$_3$ thin films

In previous studies, the strained SrCoO$_3$ samples were typically achieved via *in-situ* oxygen (or low-pressure ozone) annealing processes, in which the resulting samples usually possess pronounced oxygen vacancy contents. In this study, a two-step method has been developed to achieve the desired high quality epitaxial thin films with great oxygen stoichiometry. First, the brownmillerite SrCoO$_{2.5}$ thin films (~30 nm) were grown using the pulsed laser deposition method, and then they were post-annealed within ozone (see details in Sec. IV). The elevated annealing temperature (~300 ºC) and high ozone content (5 g/m$^3$ in 1 bar O$_2$) triggers effectively the phase transformation from brownmillerite to perovskite with excellent oxygen stoichiometry of SrCoO$_3$ [Fig. 1(a)]. To manipulate the epitaxial strain state, various substrates, (001)-oriented single crystal La$_{0.3}$Sr$_{0.7}$Al$_{0.65}$Ta$_{0.35}$O$_3$ (LSAT) (a=0.3868 nm), SrTiO$_3$ (STO) (a=0.3905 nm), and (110)$_o$-oriented DyScO$_3$ (DSO) (a$_{pseudo-cubic}$=0.3944 nm), were carefully selected, providing tensile strains of 1.0%, 2.0%, and 3.0% with respect to the lattice constant (0.3829 nm) of bulk SrCoO$_3$ [7]. As a consequence, the as-grown SrCoO$_{2.5}$ samples are epitaxially strained on all selected substrates and the crystalline orientations are perfectly correlated with that of the substrates underneath (see Supplemental Material Fig. S1 [22]). It is interesting to note that the oxygen vacancy channels of brownmillerite align along the in-plane (out-of-plane) direction with compressive (tensile) strain states, as evidenced by the presence of the superlattice peaks in the X-ray diffraction (XRD) $\theta$-$2\theta$ scans [starred peaks in Fig. 1(b)] only for the samples grown on LSAT and STO substrates with compressive strain. Despite different oxygen vacancy orientations, all samples can be nicely transformed into perovskite structure through post-deposition ozone annealing, where the $CoO_4$ tetrahedra are oxidized into the $CoO_6$ octahedra, reflecting in the XRD results as the complete absence of the superlattice peaks and the movement of the pseudo-cubic (001) and (002) diffraction peaks towards the higher angles. Furthermore, detailed X-ray reciprocal space mappings (RSM) along the pseudo-cubic (103) direction [Fig. 1(c)] were carried out to investigate the epitaxial relationship between thin films and substrates, in which the identical horizontal $q$ values between the films and substrates (marked with red dashed lines) strongly suggest that the post-ozone-annealed $SrCoO_3$ thin films are coherently strained with substrates (see Supplemental Material Tab. S1 for lattice parameters [22]). Conventionally the tensile strain in fully strained samples would introduce a systematic lattice contraction along the film normal direction. However, such trend clearly breaks down for the film grown on DSO substrate, in which a clear lattice expansion is observed. We speculate that such crystalline structure change might be related with the evolution of lattice symmetry with tensile strain. With STO and LSAT substrates, the samples would be strained into tetragonal phase from its bulk cubic compound; while with DSO substrate, the lattice structure would be largely deformed into orthorhombic phase with lower symmetry. Interesting to note that similar structural anomaly was also observed in the previous theoretical work [15], in which the lattice changes from a high symmetry *P4/mmm* to a lower symmetry *Pmc*$2_1$ at ~2% tensile strain. The high-quality epitaxial nature of $SrCoO_3$ thin films was further confirmed by the high-resolution transmission electron microscopy (HRTEM) study (see Supplemental Material Fig. S2 [22]), which ascertains that by this post-deposition ozone annealing method the $SrCoO_3$ thin films can be fully strained on the DSO substrates with as high as 3.0% strain without formation of any detectable interfacial defects that would cause strain relaxation otherwise.

As mentioned, the lack of stoichiometry of oxygen concentration in highly strained $SrCoO_3$ samples was a big challenge that researchers suffered in previous studies, where noticeable oxygen vacancies were always formed [18-20]. The presence of oxygen vacancies in thin films would strongly modify the corresponding valence states of Co ions as well as their hybridization with oxygen ions. Thus, in order to carefully evaluate the oxygen stoichiometry in our samples, we performed the valence-state-sensitive soft X-ray absorption spectroscopy (XAS) at both the Co *L*-edges and oxygen *K*-edge for all strained $SrCoO_3$ thin films. To avoid the influence of possible surface reduction, we employed a bulk sensitive fluorescence yield (FY) mode to collect the XAS signals. The XAS spectra at Co *L*-edges [Fig. 2(a)] clearly reveal that all $SrCoO_3$ films show identical features with the characteristic $Co^{4+}$ $L_3$ peak at ~780.8 eV. For reference, the $L_3$ peak of the pristine $SrCoO_{2.5}$ sample is found at ~780.3 eV with a very different line shape. This clear difference therefore suggests that the cobalt valance in all strained $SrCoO_3$ thin films is mostly 4+ and their oxygen stoichiometry difference is negligible. For comparison, previous studies reveal typically an energy shift of 0.3 eV, corresponding to the change of oxygen contents by ~0.15 per unit cell [18,20]. It is important to note that the surface sensitive total electron yield (TEY) mode also provides consistent results (see Supplemental Material Fig. S3 [22]) as compared with the spectra taken with FY mode, indicating the homogeneity nature of the films along growth direction. As the change of oxygen content can also modify the hybridization between O-*2p* and Co-*3d* orbitals, we further studied the oxygen *K*-edge XAS spectra, in which identical hybridization features at ~527.6 eV (see Supplemental Material Fig. S4 [22]) further prove the excellent oxygen stoichiometry, since this hybridization peak in oxygen deficient $SrCoO_{2.5}$ would be dramatically suppressed and shifted toward higher energy (~529.2 eV) [23,24]. With all these careful analyses, we can confirm that the post-deposition ozone annealing method achieves coherently-strained and stoichiometric SrCoO$_3$ thin films.

## B. Robust ferromagnetism in the highly strained samples

With these high-quality samples, we further investigated their intrinsic magnetic properties. Since the large paramagnetic background of DSO prohibits the direct magnetic measurement of the thin films grown on it, we carried out the element-specific and magnetic-resolved soft X-ray magnetic circular dichroism (XMCD) measurement [25]. Fig. 2(b) shows the measured XMCD spectra of Co $L$-edges at 20 K, in which the pronounced (~18% at $L_3$ edge) XMCD signals with similar characteristic peaks manifest that all strained SrCoO$_3$ thin films possess robust magnetizations. Fig. 2(c) represents the magnetic field dependent XMCD for all three samples, which shows well-defined hysteresis loops, confirming the FM ground states. Using the XMCD sum rule (see Ref. [25] and Supplemental Material Fig. S5 [22]) we calculated the spin and orbital contributions for the total magnetization as around ~1.0 $\mu_B$/Co and ~0.4 $\mu_B$/Co, respectively. This result is consistent with the experimentally reported value for bulk SrCoO$_3$ [7] as well as slightly strained (1.0%) thin films [23,24], however, it is in disagreement with some previous theoretical calculations that predicted an AFM ground state for highly strained samples [15,16]. To further investigate the FM state, we carried out temperature-dependent XMCD measurement for all strained samples (see Supplemental Material Fig. S6 [22]) with the calculated total magnetizations summarized in Fig. 2(d), which are consistent with the macroscopically measured magnetizations. The magnetic transition temperatures are estimated to be ~250 K, which is consistent with the previously reported value in high quality SrCoO$_3$ samples grown on LSAT [23,24,26].

## C. Strain induced metal to insulator transition

While showing similar and robust FM states, this series of strained SrCoO$_3$ samples exhibit very different electronic properties, as revealed by the temperature dependent electrical transport measurements shown in Fig. 3(a). The SrCoO$_3$ sample grown on LSAT substrate shows a good metallic characteristic with residual resistivity of ~$10^{-4}$

Ω·cm at ~15 K, which is much lower than previously reported values of single crystalline bulk [7] and thin films [24], however consistent with our recent studies of SrCoO$_3$ thin films obtained with ionic liquid gating [26]. This fact again confirms that our ozone-annealed SrCoO$_3$ samples are of high crystalline quality with minimal oxygen vacancy concentration. This temperature dependent curve follows nicely the $T^2$ relationship, in accordance with the Fermi liquid behavior of this metallic sample. Interestingly, with the increase of tensile stain, the conductivity is dramatically suppressed and a distinct insulating behavior is developed in the sample grown on DSO. Specifically, we have observed a more than two orders of magnitude enhancement of resistivity at room temperature between the LSAT and DSO samples. A careful analysis reveals that the temperature dependence of such highly strained sample can be fitted with the three dimensional Mott variable range hopping model [27,28] with $\rho(T) = \rho_0 e^{(\frac{T_0}{T})^{1/4}}$. For SrCoO$_3$ grown on STO with tensile strain at the intermediate state, the resistivity remains insensitive across a wide temperature range [blue line in Fig. 3(a)], indicating that the SrCoO$_3$ film grown on STO is close to the phase boundary of metal insulator transition. For comparison, SrCoO$_{2.5}$ thin film grown on LSAT substrate shows a dramatically reduced room temperature conductivity as compared with all SrCoO$_3$ thin films, and its exponential temperature dependence follows nicely the thermal activated semiconducting behavior.

To provide further insights of the strain induced metal to insulator transition, we measured the optical conductivity spectra for selected SrCoO$_3$ samples as shown in Fig. 3(b). A Drude-like feature is observed for the sample grown on LSAT substrate, which is consistent with its metallic nature. To investigate the insulating state, we replaced DSO substrate with TbScO$_3$ (TSO) ($a_{pseudo-cubic}$=3.953 nm) with the strain of 3.2% to avoid the absorption background from DSO substrate. With this sample, the Drude feature is significantly suppressed with a spectra weight transferred to higher photon energy, which clearly indicates an insulating state of the SrCoO$_3$ with large strain. As a control experiment, reference spectra taken from a typical Mott insulator LaCoO$_3$ (with Co$^{3+}$), show a further shift of spectra weight toward higher photon energy with negligible spectra intensity at low energy. The notable spectral weight transfer over the large energy scale (~2 eV) between these samples at different strain states indicates that SrCoO$_3$ is on the verge of the electron-correlation induced metal to insulator transition or the Mott transition, critically driven by lattice strains [29].

With these evidences, we demonstrated experimentally a persistent FM state in highly strained SrCoO$_3$ thin films with a metal to insulator transition occurring at ~2.0% tensile strain state. In previous theoretical works, the FM state in SrCoO$_3$ was usually attributed to the ferromagnetically coupled intermediate-spin Co $d^6L$ through Zener double-exchange mechanism [10,11], in which the FM ground state would be accompanied by a metallic electric state. This would lead to an immediate question that how the highly strained SrCoO$_3$ thin films could maintain the FM ground state while undergoing a metal to insulator transition.

### D. Electronic structure calculations

In order to understand the underlying mechanism, we performed DFT calculations as well as DFT+DMFT calculations. As shown in Fig. 4(a), our spin-resolved density of states (DOS) for relaxed SrCoO$_3$ is quantitatively consistent with the previous calculated result [13]. In contrast to the nearly fully-occupied majority spin states, minority spin states are partially occupied and make primary contribution to the electric conductance, in which the Fermi level crosses mainly at the Co $t_{2g}$ states for relaxed SrCoO$_3$. The orbital occupancy of the fully relaxed sample obtained with our DFT+DMFT is indeed qualitatively in agreement with our DFT results (see Supplemental Material Figs. S7(a) and S7(b) [22]). On the other hand, in tensile strained samples, the orbital degeneracy in $t_{2g}$ states ($xz$, $yz$ vs. $xy$) and in $e_g$ states ($x^2-y^2$ vs. $3z^2-r^2$) is expected to be lifted. Indeed, both DFT and DFT+DMFT results show such degeneracy lifting (see Supplemental Material Figs. S7(a) and S7(b) [22]), while the total Co-3$d$ occupation and the ordered Co moments are only weakly dependent on the strain (see Supplemental Material Figs. S7(c) and S7(d) [22]). The degeneracy lifting is much larger in the DFT+DMFT calculation, giving a stronger strain dependent orbital polarization. Under 4% tensile strain, the orbital splitting is so strong in the minority spin states that the Fermi level crosses primarily at *xz* and *yz* bands in DFT+DMFT data, while in DFT data the Fermi level crosses at *xz*, *yz* and *xy* bands nearly equally (see Supplemental Material Figs. S8(a)-S8(d) for comparison [22]), i.e., orbital polarization is amplified by correlation effects [30-32]. This would result in one electron occupation in the twofold degenerate minority spin *xz* and *yz* orbitals, and potentially lead to the cooperative Jahn-Teller effect accompanied by alternating spatial arrangement of *xz* and *yz* orbitals. To simulate this situation, we introduced a tiny potential splitting between the *xz* and *yz* orbitals, $\Delta \varepsilon = 0.1$ eV, and carried out DFT+DMFT calculation. As shown in Fig. 4(b), the tiny splitting is enough to separate the *xz* and *yz* bands at the Fermi level as marked by arrows. The difference in the transport property could be explained more clearly in momentum resolved spectral functions as shown in Figs. 4(c) and 4(d). The spectral intensity at the Fermi level is suppressed in the strained sample with such a small perturbation (0.1 eV), which could be regarded as a precursor of the electron localization. However, even with such a splitting inside minority bands, the FM ordering remains stable, and the ordered moment is barely changed.

Here, in contrast to the metallic state at unstrained or low strained $SrCoO_3$ with the *xz/yz* degeneracy, where the FM interaction is mediated by the double-exchange interaction due to mobile minority electrons [Fig. 5(a)], the super-exchange interaction with the alternating *xz/yz* ordering is responsible for the FM coupling due to the Kugel-Khomskii mechanism [33] as schematically shown in Fig. 5(b). We note that a similar FM interaction is realized in lightly-doped CMR manganite [34]. This picture strongly contrasts with the previously predicted AFM ground state by first principles calculations [15]. Indeed, using DFT, we confirmed that AFM ordering is also more stable than FM ordering for 4% strained sample by 0.36 eV per formula unit, which is accompanied by the change of the electron occupation $N$ and the spin moment $M$ on Co-3*d* states from $N \approx 6.8$ and $M \approx 2.9\ \mu_B$ at FM to $N \approx 6.7$ and $M \approx 3.1\ \mu_B$ at AFM in DFT (both are close to IS state at $d^{6.5\sim7}$). These differences become even larger in our DFT+DMFT calculations; $N \approx 6.4$ and $M \approx 2.7\ \mu_B$ at FM (IS state at $d^{6.5}$) and $N \approx 6$ and $M \approx 3.7\ \mu_B$ at AFM (HS state at $d^6$). This drastic change of $N$ and $M$ comes from the fact that intermediate-spin states, in which both $t_{2g}$ and $e_g$ orbitals are partially filled, are unstable compared with low-spin and high-spin states in the atomic limit [35]. This energetics might be overlooked in the DFT calculations, and the comparison of the total energy between different spin states as well as the examination of *xz/yz* orbital ordering within the DFT+DMFT framework are desirable but left for future studies.

## III. SUMMARY

To summarize, this work demonstrated a robust FM state in SrCoO$_3$ thin films under tensile strain beyond the critical strain value above which a transition from FM to AFM was predicted. We found the coexistence of the ferromagnetism and insulating transport behavior in highly strained samples, indicating a different mechanism being responsible for the FM ordering in the strained thin films rather than the Zener's double-exchange mechanism. Based on the DFT+DMFT calculations indicating strong orbital polarization under strain, we propose orbital ordering to induce the FM super-exchange interaction with the splitting inside minority bands, which might finally cause a full gap. Furthermore, these results clearly highlight the importance of excellent oxygen stoichiometry for the studies of strained engineered complex oxides.

## IV. METHODS

### A. Synthesis of high quality SrCoO$_3$ thin films

Brownmillerite SrCoO$_{2.5}$ thin films were grown by a customized reflection high-energy electron diffraction (RHEED) assisted pulsed laser deposition system, at a growth temperature of 750 ºC and oxygen pressure of 0.125 mbar. The laser energy (KrF, $\lambda$ = 248 nm) was fixed at 1.1 J/cm$^2$ with the repetition rate of 2 Hz. After the growth, the samples were cooled down to room temperature at a cooling rate of 10 ºC/min with 0.125 mbar oxygen to achieve high quality epitaxial SrCoO$_{2.5}$ thin films with the thickness of ~30 nm. The sample thickness was confirmed with the X-ray reflectometry measurements, and the crystalline structures of thin films were characterized by a high-resolution four-circle X-ray diffractometer (Smartlab, Rigaku). The films were then post-annealed in a chamber filled with 1 bar mixed ozone (5 g/m$^3$) and oxygen with the flow rate of 0.5 L/min at 300 ºC for 40 mins, which nicely triggers the phase transformation from brownmillerite SrCoO$_{2.5}$ into perovskite SrCoO$_3$.

### B. High-resolution scanning transmission electron microscopy characterization

The atomic-scale crystalline structures of the highly strained SrCoO$_3$ thin films were characterized with a high-resolution scanning transmission electron microscope (STEM, ARM-200CF, JEOL) operated at 200 keV and equipped with double spherical aberration (Cs) correctors.

### C. Electrical transport and magnetic property measurements

A Physical Property Measurement System (PPMS, Quantum Design) with lock-in amplifiers (Model SR830 DSP, Stanford research systems) was employed to measure the temperature dependent resistivity, in which four-probe method was employed to eliminate the contact resistance. Temperature dependent magnetization was measured using a Magnetic Property Measurement System (MPMS 3, Quantum Design) with the temperature ramping up from 4 K to 300 K after a 2 T field cooling. In order to avoid the influence of the strong paramagnetic signal from DyScO$_3$ substrate, we measured the remnant magnetization with zero-field, while for other measurements, the magnetic field of 2 T was applied during the measurements.

### D. Optical conductivity measurement

The reflectivity spectra were measured at room temperature with a spectrometer (Bruker IFS66v, Jasco MSV370), which covers the visible and infrared range with the energy from 0.3 eV to 3 eV. For this measurement, thin films were prepared on double-side polished substrates. To obtain enough optical intensity, we used slightly thicker samples with the thickness of ~50 nm as comparing with other measurements. The optical conductivity spectra were derived by the Kramers-Kronig analysis on the basis of the two-layer model.

### E. X-ray absorption measurements

Soft X-ray absorption spectra (XAS) at Co-*L* edges and O-*K* edge were carried out at Beamlines 4.0.2 and 6.3.1 of Advanced Light Source with both total electron yield (TEY) mode and luminescence yield (LY) mode and then confirmed at Beamline I06 of Diamond Light source with TEY and fluorescence yield (FY) modes and at BL25SU of SPring-8 with only TEY mode. The XAS spectra were normalized to the photon flux measured with the photocurrent from a clean gold mesh. For the X-ray magnetic circular dichroism (XMCD) measurements, the incident angle was fixed at 60 degrees, the magnetic field (2 T) was applied along the beam incident direction, and circularly polarized X-ray was employed. The XAS spectra presented in this work were taken at 300 K, while the XMCD spectra were taken from 3 K to 300 K. For the TEY mode, the signal was obtained by collecting the total electron yield current from the sample, which is therefore a surface sensitive technique. While for the FY (LY) mode, the signal was a collection of the fluorescence (luminescence) intensity from the whole sample, and therefore is bulk sensitive. It is worth noting that due to the self-absorption effect of FY, we observed clear different $L_2/L_3$ peak ratios between FY and TEY modes.

### F. Electronic structure calculations

We carry out density functional theory (DFT) calculations using Vienna *ab initio* Simulation Package (VASP) [36] with the projector augmented wave method [37] and the Perdew–Burke–Ernzerhof (PBE) exchange-correlation functional [38]. We used for Co and O, standard potentials (Co and O in the VASP distribution) and for Sr a potential in which *s* states are treated as valence states ($Sr_{sv}$). Structural optimization was performed using the Dudarev type [39] on-site effective Coulomb interaction on Co *d* states with $U_{eff}$ = U-J = 1.5 eV (U = 2.5 eV and J = 1.0 eV) [16] with the FM spin polarization. A kinetic energy cut-off of 600 eV and a Monkhorst-Pack *k*-mesh of 12×12×12 are used for the plane-wave expansion of wavefunctions and *k*-space integration, respectively. The calculated equilibrium lattice constants are 3.834 Å, which is very close to the experimental value of 3.829 Å [7]. For strain-free cubic perovskite $SrCoO_3$, all the lattice parameters (*a*, *b*, *c*) are optimized, and for strained

SrCoO$_3$ the *c*-axis lattice constant is optimized under *ab*-plane biaxial tensile strain of 1% to 4%.

To extracting the parameters to describe the hopping between Co *d* and/or O *p* orbitals, maximally localized Wannier functions were further constructed with the Wannier90 code [40] using non-spin-polarized PBE-only wave functions calculated on the above optimized structures. Fourteen Wannier functions were thus obtained corresponding to the five *d* orbitals on one Co cation and the nine *p* orbitals on three equivalent O anions. The spreads of the Wannier functions were all smaller than 1 Å$^2$.

For comparison with DFT+DMFT results, additional DFT calculations are carried out using the optimized structure as mentioned above. On-site effective Coulomb interaction on Co *d* states is taken as U$_{eff}$ = 5.4 eV [41], which gives the better agreement with the experiment and DFT+DMFT results in terms of ordered Co moments. Orbital resolved Co *d* charge density and magnetization are computed from the local density of states projected onto Co *d* shells within the Wigner Seitz radius of 1.302 Å.

DMFT calculations were performed using an effective model expressed as $H = H_{band} + \sum_{i \in \text{Co } d} H_{int,i}$. The band part $H_{band}$ is given by

$$H_{band} = \sum_{\vec{k}\sigma} [\hat{d}^\dagger_{\vec{k}\sigma}, \hat{p}^\dagger_{\vec{k}\sigma}] \begin{bmatrix} \hat{\varepsilon}^{dd}_{\vec{k}} - \varepsilon_{DC}\hat{1} & \hat{\varepsilon}^{dp}_{\vec{k}} \\ \hat{\varepsilon}^{pd}_{\vec{k}} & \hat{\varepsilon}^{pp}_{\vec{k}} \end{bmatrix} \begin{bmatrix} \hat{d}_{\vec{k}\sigma} \\ \hat{p}_{\vec{k}\sigma} \end{bmatrix}.$$

Here, $\hat{d}_{\vec{k}\sigma}(\hat{p}_{\vec{k}\sigma})$ is a vector consisting of annihilation operators of Co *d* (O *p*) electrons with spin $\sigma$ and momentum $\vec{k}$. Matrices $\hat{\varepsilon}^{\alpha\beta}_{\vec{k}}$ are the momentum dependent hybridization functions in the Wannier basis obtained in the DFT calculations. $\alpha$ and $\beta$ run through Co *d* and O *p* states. $\varepsilon_{DC}$ is so-called double-counting correction, which subtracts Hartree-type Coulomb repulsion effects already included within DFT calculations. The interaction part $H_{int,i}$ is formally expressed as $H_{int,i} = \sum_{\alpha,\beta,\gamma,\delta} U_{\alpha\beta\gamma\delta} d^\dagger_{i\alpha} d^\dagger_{i\beta} d_{i\delta} d_{i\gamma}$ with $\alpha, \beta, \gamma, \delta$ indexing both spin and orbital on the Co *d* shell. Coulomb interaction $U_{\alpha\beta\gamma\delta}$ is expressed in terms of Slater integrals $F^{0,2,4}$. Taking the screened Coulomb interaction $U(= F^0) = 10$ eV and exchange interaction

$J = 1$ eV [42] with $F^4/F^2 = 5/8$ [43], we determine $F^2 = 8.615$ eV and $F^4 = 5.385$ eV. The DMFT procedure maps this interacting lattice model into an interacting impurity model consisting of interacting Co $d$ states and the effective hybridization that will be solved self-consistently [44]. To solve the impurity model, we use the hybridization-expansion version of the continuous-time quantum Monte-Carlo method [45] at $T = 0.02$ eV = 232 K. We keep only density-density components of $U_{\alpha\beta\gamma\delta}$ so that the efficient segment algorithm is utilized [46].

For the double-counting correction $\varepsilon_{DC}$, we used the procedure proposed in Ref. [47]:

$$\varepsilon_{DC} = \frac{1}{N_d} Re \sum_\alpha \Sigma_\alpha(i\omega_n \to i\infty)$$

Here, $N_d$ is the number of Co $d$ orbital times 2 (spin), i.e., 10. $\Sigma_\alpha(i\omega_n)$ is the Co $d$ electron self-energy as a function of fermionic Matsubara frequency $\omega_n = (2n + 1)\pi T$. Using this formula, $\varepsilon_{DC}$ is updated at each DMFT iteration. This method has been successfully applied to describe metallic systems [47]. The electron spectral functions are computed using the self-energy that is analytically continued to the real frequency axis by the maximum entropy method [48].

## ACKNOWLEDGEMENTS

This study was financially supported by the Basic Science Center Project of NFSC under grant No. 51788104; the National Natural Science Foundation of China (grant No. 51872155); the National Basic Research Program of China of grant No. 2016YFA0301004; the Beijing Advanced Innovation Center for Future Chip (ICFC); and Engineering and Physical Sciences Research Council with grant reference EP/N016718/1. This research used resources of the Advanced Light Source, which is a DOE Office of Science User Facility under contract no. DE-AC02-05CH11231. The research at ORNL was supported by US DOE, Office of Science, Basic Energy Sciences, Materials Sciences and Engineering Division. N. P. Lu acknowledges support from the National Natural Science Foundation of China (grant No. 11974401), the Hundred Talents Program of Chinese Academy of Science of China, and the Strategic

**Captions**

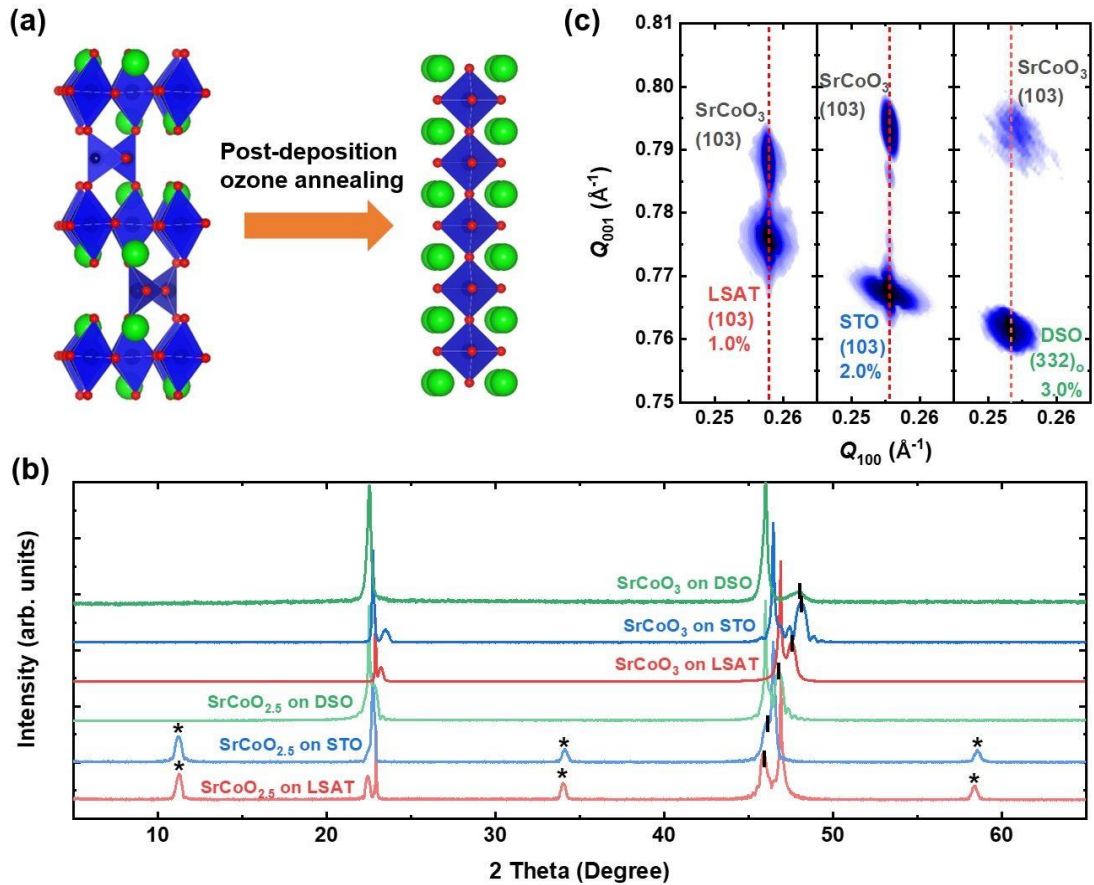

**FIG. 1. Strain engineered SrCoO₃ thin films.** (a) Schematic diagram of post-deposition ozone annealing induced phase transformation from $SrCoO_{2.5}$ to $SrCoO_3$. (b) X-ray diffraction $\theta$–$2\theta$ scans of pristine $SrCoO_{2.5}$ and post-deposition ozone annealed $SrCoO_3$ thin films grown on LSAT (001), STO (001) and DSO (110)$_o$ substrates (with 1.0%, 2.0% and 3.0% tensile strain, respectively). Superlattice peaks with star marks suggest that the oxygen vacancy channel prefers in-plane orientation for the films grown on LSAT and STO substrates, while with post-deposition ozone annealing process the superlattice peaks are vanished for all films. (c) Reciprocal space mapping along the pseudo-cubic (103) direction of $SrCoO_3$ thin films. The perfect alignment of the $Q_{100}$ values between the thin films and the substrates is highlighted by the red dashed lines, suggesting that the films are coherently strained by all the substrates.

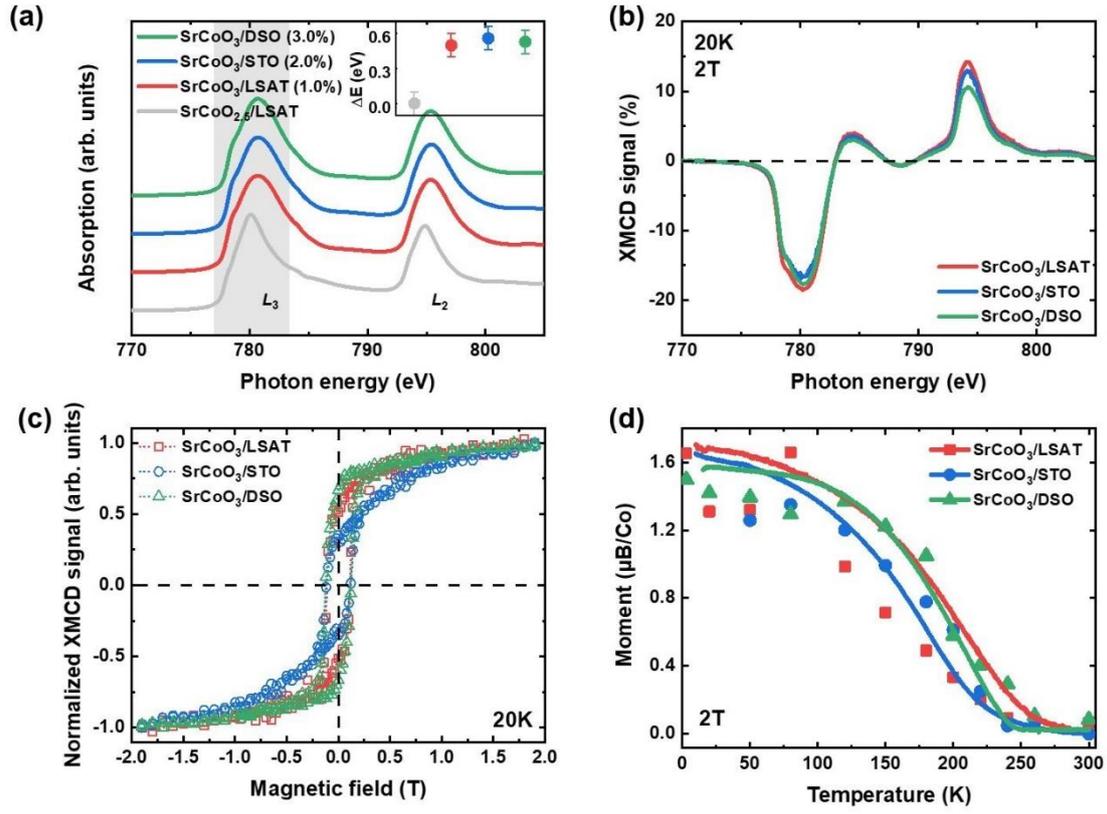

**FIG. 2. Robust ferromagnetism in strained SrCoO$_3$ thin films.** (a) Comparison of the Co *L*-edges X-ray absorption spectra taken using the fluorescence yield mode. The inset shows negligible peak shift (within the resolution of 0.1 eV) at Co $L_3$-edge for all SrCoO$_3$ samples. (b) Co *L*-edges XMCD spectra taken from strain engineered SrCoO$_3$ thin films at 2 T and 20 K. (c) Magnetic field dependent XMCD signals for Co $L_3$-edge in strained SrCoO$_3$ thin films measured at 20 K. (d) Summary of the temperature dependent magnetizations calculated by XMCD sum rule (solid squares, circles and triangles) and macroscopic magnetizations measured by SQUID (solid lines) (see Sec. IV) for SrCoO$_3$ thin films at various strain states.

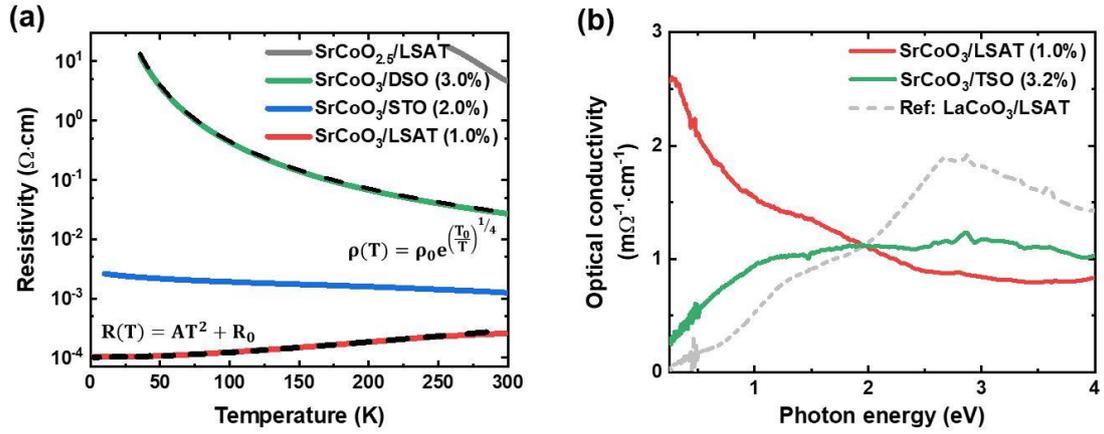

**FIG. 3. Strain induced metal to insulating transition in SrCoO₃ thin films.** (a) Temperature dependent resistivity of strained SrCoO$_3$ thin films grown on various substrates. The result of SrCoO$_{2.5}$ grown on LSAT is shown in grey as reference. Fitting of the resistivity results are shown in dotted lines. (b) The room-temperature optical conductivity spectra from SrCoO$_3$ thin films grown on LSAT (1.0%) and TSO (3.2%) substrates (we replaced the DSO substrate with TSO to avoid the absorption background from the DSO substrate.). The gray dotted line shows the result from a typically Mott insulator of LaCoO$_3$ (grown on LSAT substrate) for comparison purpose.

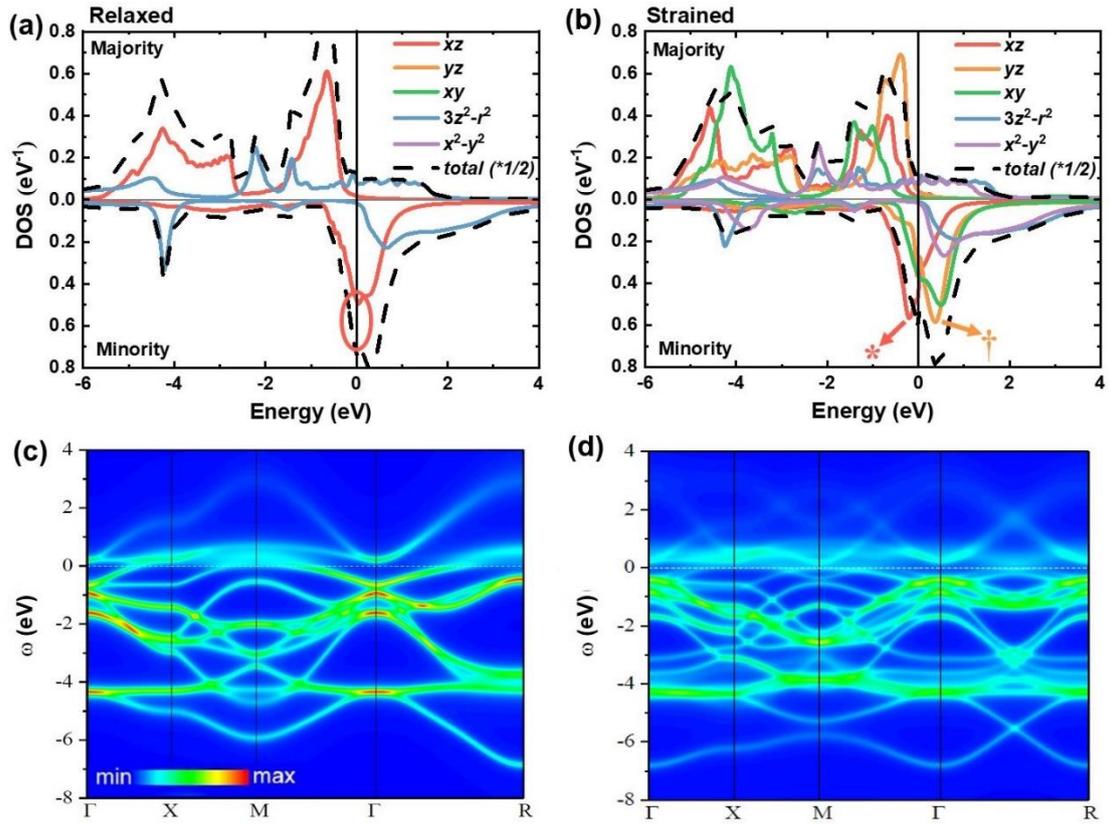

**FIG. 4. Evolution of the electronic band structures for strained SrCoO3 thin films with DFT+DMFT.** Spin-resolved projected density of states (DOS) spectra for (a) the relaxed SrCoO$_3$ and (b) the strained SrCoO$_3$ (4% with $xz/yz$ alternating splitting $\Delta\varepsilon = 0.1$ eV). The arrows highlight a clear splitting inside minority bands (∗ for minority $xz$ band and † for minority $yz$ band) at the Fermi level for the strained SrCoO$_3$ as compared with non-splitting in relaxed SrCoO$_3$ marked by a red circle. The suppression of the spectral weight at the Fermi level is more clearly seen by comparing the electronic band dispersion for (c) the relaxed and (d) the strained SrCoO$_3$.

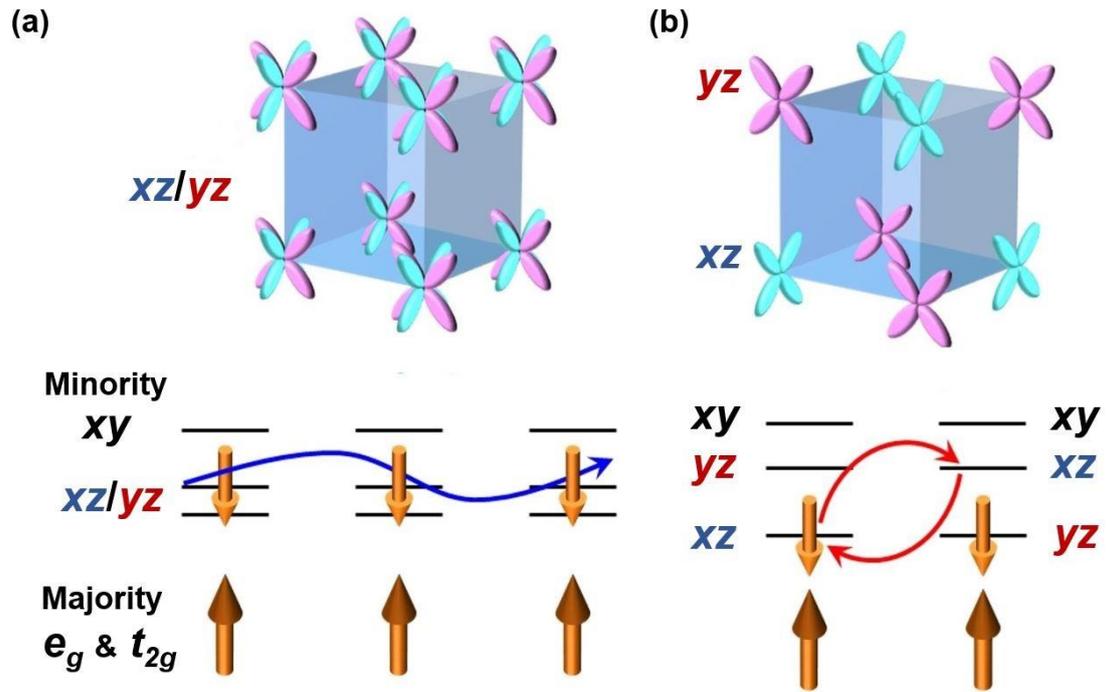

**FIG. 5. Schematics of orbital structure and its correlation with ferromagnetic ground state.** (a) Unstrained (or low strained) $SrCoO_3$ with twofold degenerate *xz* and *yz* bands, and (b) highly strained $SrCoO_3$ with alternating *xz*/*yz* orbital ordering. In (a), mobile minority electrons in the *xz*/*yz* band mediate double-exchange FM interaction. In (b), on the other hand, the virtual excitation of minority electrons from occupied *xz* (*yz*) orbital to neighboring unoccupied *xz* (*yz*) orbital induces Kugel-Khomskii-type FM super-exchange interaction.

**Supplementary Material**

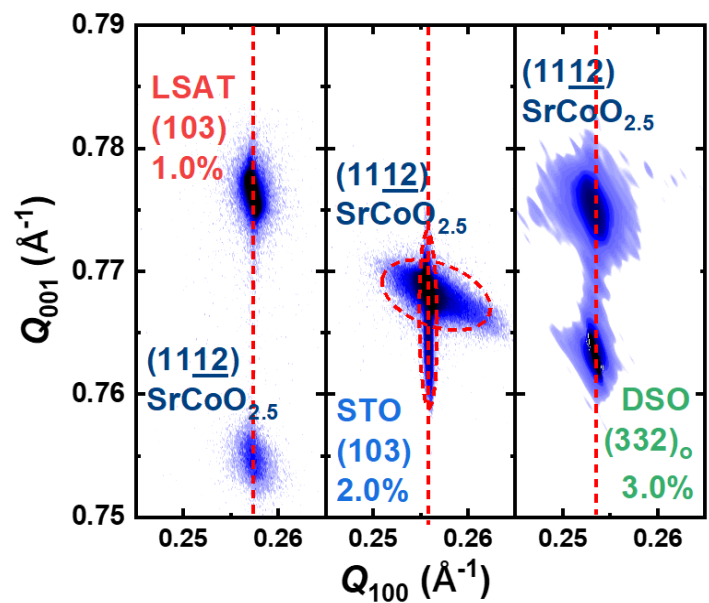

**Fig. S1.** Comparison of reciprocal space mapping (RSM) results of the (11$\underline{2}$) plane in SrCoO$_{2.5}$ thin films grown on LSAT (001), STO (001) and DSO (110)$_o$ substrates. The perfect alignment of $Q_{100}$ values between thin films and substrates is highlighted by the red dashed lines, suggesting that the films are coherently strained by all substrates.

| Substrate | In-plane lattice (nm) | C-lattice constant (nm) | Strain |
|:---:|:---:|:---:|:---:|
| LSAT | 0.387 | 0.381 | 1.0% |
| STO | 0.391 | 0.378 | 2.0% |
| DSO | 0.394 | 0.378 | 3.0% |

**Tab. S1**. Summary of SrCoO$_3$ thin films lattice parameters extracted from the RSM measurements.

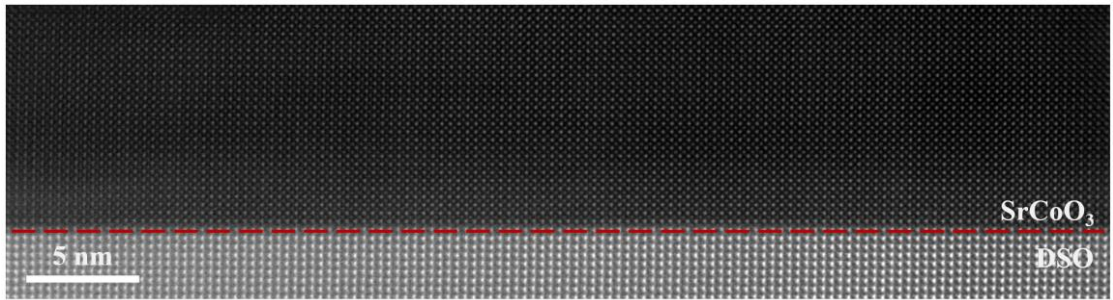

**Fig. S2**. High-resolution transmission electron microscopy images of highly strained SrCoO$_3$ thin film grown on DSO (110)$_o$ substrate. The zone axis is along the (010)$_{pc}$ direction.

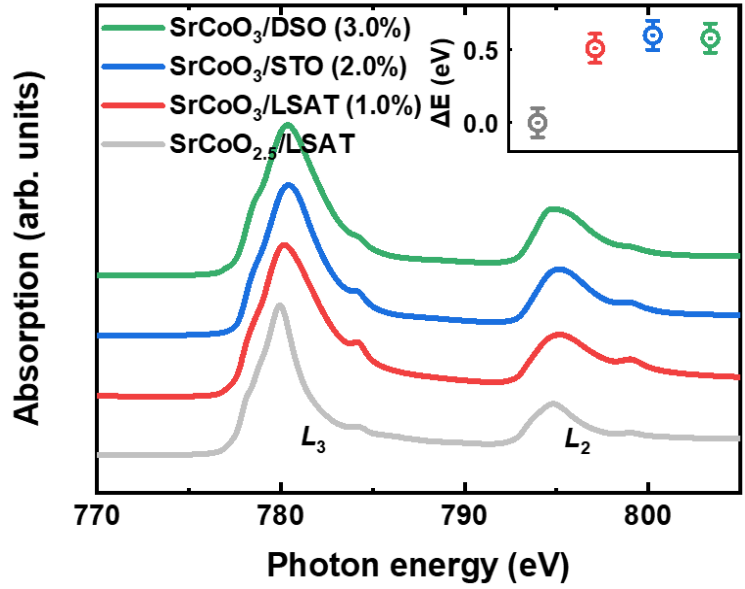

**Fig. S3.** Co *L*-edges XAS spectra of strained SrCoO$_3$ taken with total electron yield mode. The result from SrCoO$_{2.5}$ is shown as reference with gray line. The inset shows negligible peak shift (within the resolution of 0.1 eV) at Co *L*$_3$-edge.

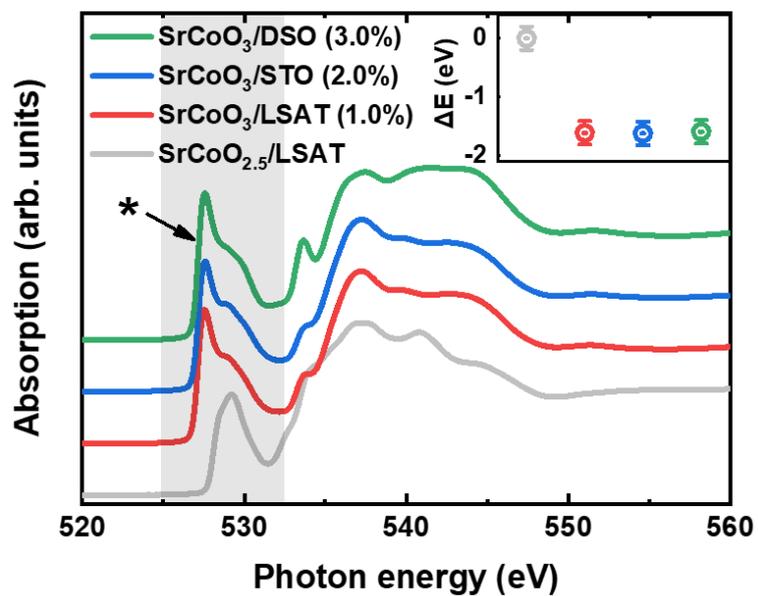

**Fig. S4.** Comparison of O *K*-edge XAS spectra (with TEY mode) taken from strain SrCoO$_3$ thin films at various strain states. The spectra taken from SrCoO$_{2.5}$ is shown as reference with gray line. The inset indicates clearly negligible shift of the hybridization peak (within the resolution of 0.1 eV) among three films, as labeled by the star marker.

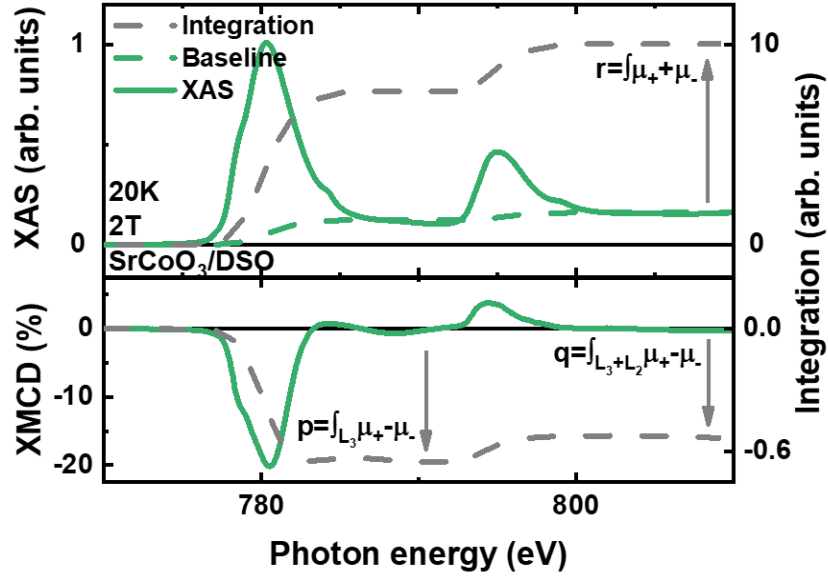

**Fig. S5.** Representative magnetic sum-rule calculation from the XMCD spectra. The data were obtained from a SrCoO$_3$ thin film grown on DSO substrate measured at 20 K with the magnetic field of 2 T. After deducting a baseline of piecewise function, the integrations of $\mu_+ + \mu_-$ for $L_{3,2}$, $\mu_+ - \mu_-$ for $L_3$ and $\mu_+ - \mu_-$ for $L_{3,2}$ were calculated as $r$, $p$ and $q$ respectively. And then, the orbital angular and spin angular momentum can be calculated as $m_{orb} = -\frac{4q(10-n_{3d})}{3r}$ and $m_{spin} = -\frac{(6p-4q)(10-n_{3d})}{r}$, in which $n_{3d}$ is the nominal number of $d$ electrons (5 for Co$^{4+}$ here).

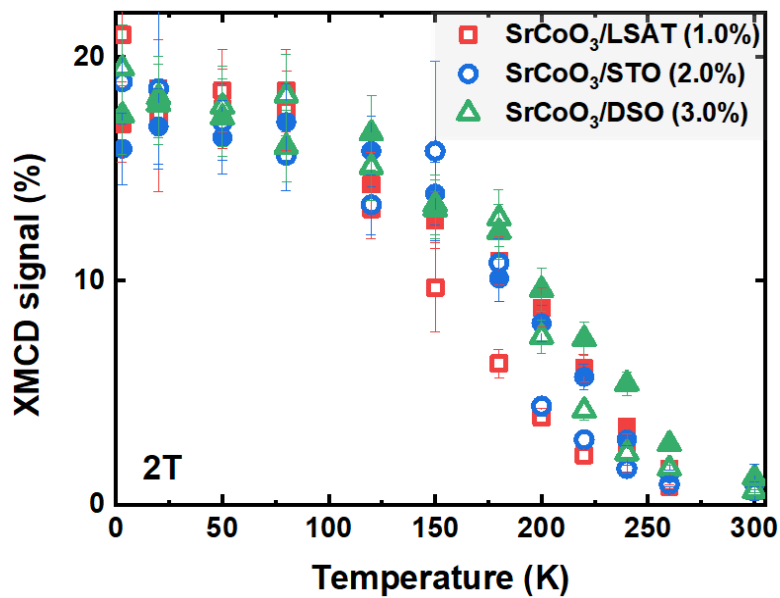

**Fig. S6**. Temperature dependent XMCD spectral intensities (hollow graphics for TEY mode and solid graphics for FY mode) for epitaxial SrCoO$_3$ thin films with various strain states.

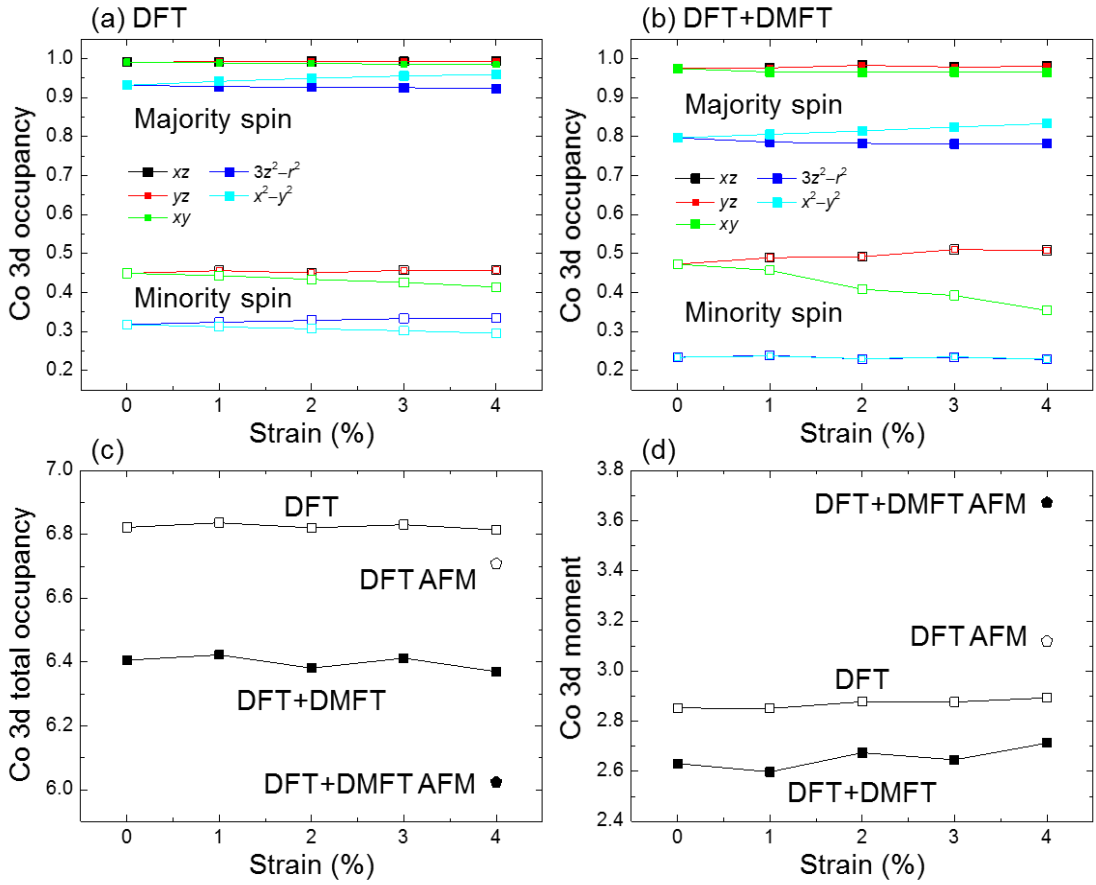

**Fig. S7**. Comparison between DFT and DFT+DMFT calculated results. The strain dependence of Co-3$d$ occupancy computed by DFT (a) and DFT+DMFT (b) in FM states. The strain dependence of Co-3$d$ total occupancy (c) and Co-3$d$ spin moment (d) in FM states. In (c) and (d), results for AFM states are also shown (in pentagons) for reference.

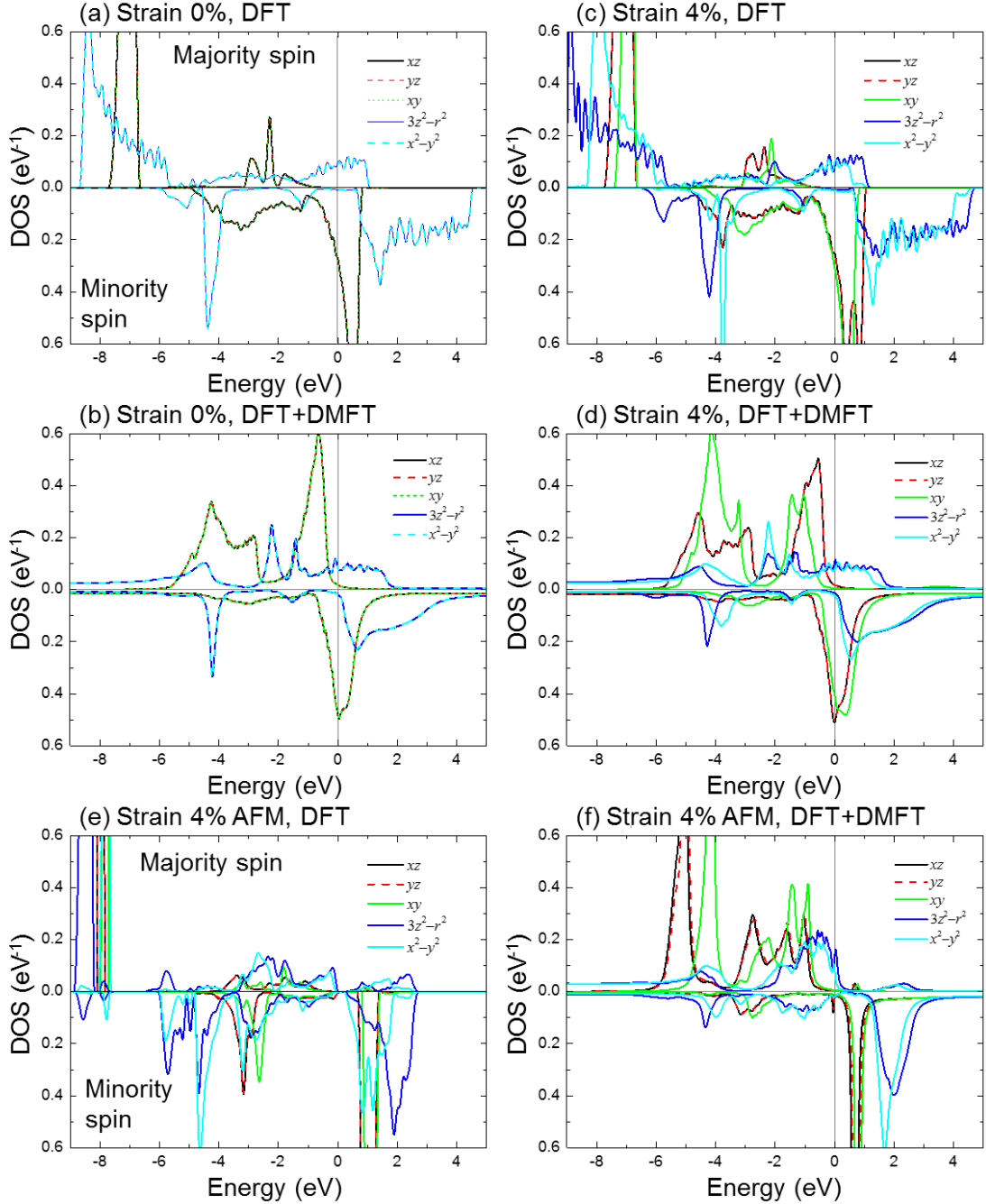

**Fig. S8**. Comparison of calculated density of states between DFT and DFT+DMFT methods. Co-3*d* partial DOS at 0% strain by DFT (a) and DFT+DMFT (b), and at 4% strain by DFT (c) and DFT+DMFT (d) both in FM states. Co-3*d* partial DOS at 4% strain by DFT (e) and DFT+DMFT (f) in AFM states. The high-spin state is realized in DFT+DMFT for AFM phase (f), while for other cases intermediate-spin states are realized (a)-(e).